\documentclass[aps,prl,reprint,groupedaddress]{revtex4-1}

\usepackage{graphicx}
\usepackage{dcolumn}
\usepackage{bm}
\usepackage{amsmath}

\begin{document}

\title{Random death process for the regularization of subdiffusive anomalous equations}
\author{Sergei Fedotov}
\author{Steven Falconer}
\date{\today }

\begin{abstract}
Subdiffusive fractional equations are not structurally stable with respect
to spatial perturbations to the anomalous exponent (Phys. Rev. E 85, 031132 (2012)). The question arises of
applicability of these fractional equations to model real world phenomena. To
rectify this problem we propose the inclusion of the random death process
into the random walk scheme from which we arrive at the modified fractional master
equation. We analyze the asymptotic behavior of this equation, both
analytically and by Monte Carlo simulation, and show that this equation is
structurally stable against spatial variations of anomalous exponent.
Additionally, in the continuous and long time limit we arrived at an unusual
advection-diffusion equation, where advection and diffusion coefficients
depend on both the death rate and anomalous exponent. We apply the
regularized fractional master equation to the problem of morphogen gradient
formation.
\end{abstract}

\maketitle

\affiliation{School of Mathematics, The University of Manchester, Manchester
M60 1QD, United Kingdom}

\affiliation{School of Mathematics, The University of Manchester, Manchester
M60 1QD, United Kingdom}

Anomalous subdiffusion, where the mean squared displacement grows
sub-linearly with time $\langle x^{2}(t)\rangle \sim t^{\nu },$ where the
anomalous exponent $\nu <1$, is an observed natural phenomena \cite%
{klages2008anomalous}. It is seen in areas as varied as dispersive charge
transport in semi-conductors \cite{Scher1975}, ion movement in spiny
dendrites \cite{Santamaria2006,*Fedotov2008,*Fedotov2010_2}, protein
transport on cell membranes \cite{Ghosh1994,*Feder1996}. In the classical
paper \cite{Metzler1999_2}, Metzler, Barkai, and Klafter introduced the
fractional Fokker-Planck equation (FFPE) that describes anomalous
subdiffusion of particles in an external field, $F(x)$. This equation for
the probability density $p(x,t)$ is written as $\frac{\partial p}{\partial t}%
=\mathcal{D}_{t}^{1-\nu }L_{FP}p(x,t)$, where $L_{FP}=K_{\nu }\left[ \frac{%
\partial ^{2}}{\partial x^{2}}-\frac{\partial }{\partial x}\frac{F(x)}{k_{B}T%
}\right] $ is the Fokker-Planck operator, $K_{\nu }$ is the anomalous
diffusion coefficient and $\mathcal{D}_{t}^{1-\nu }$ is the
Riemann-Liouville fractional derivative of order $1-\nu $, defined as
\begin{equation}
\mathcal{D}_{t}^{1-\nu }p(x,t)=\frac{1}{\Gamma (\nu )}\frac{\partial }{%
\partial \tau }\int_{0}^{t}\frac{p(x,\tau )d\tau }{(t-\tau )^{1-\nu }}.
\label{RL}
\end{equation}%
It was shown that the external field $F(x)$ leads to a stationary solution
to the FFPE in the form of the Boltzmann distribution \cite{Metzler2000c}. However, in a recent paper \cite%
{Fedotov2012}, we have demonstrated that this fundamental result is not
structurally stable with respect to spatial variations of the anomalous
exponent $\nu $. Small, non-homogeneous in space, variations of $\nu $
destroy the stationary solution to the FFPE. In fact, even the simple
one-dimensional fractional subdiffusion equation with constant anomalous
exponent and $F(x)=0$, in the finite domain $[0,L]$ with reflective boundary
conditions, is structurally unstable. This equation should give a uniform
stationary distribution over the interval $[0,L]$ in the long-time limit.
However, if we use slightly non-uniform anomalous exponent $\nu (x)$, the
probability density $p(x,t)$ will be completely different from the uniform
distribution: as $t\rightarrow \infty $ it concentrates at the point where $%
\nu (x)$ has a global minimum on $[0,L]$. We called this phenomenon
anomalous aggregation \cite{Fedotov2011}.  Since it is impossible to have a
completely homogeneous environment, in which $\nu $ is uniform, the question
arises as to whether the fractional equations with constant anomalous
exponents are useful models for any real phenomena involving subdiffusion.
This question is of great importance for the problem of a morphological
patterning of embryonic cells, which is controlled by the distribution of
signaling molecules known as morphogens \cite{Rogers2011,Hornung,Yuste2010}.
To ensure robust pattern formation, the morphogen gradients must be
structurally stable with respect to the spatial variations of environmental
parameters including the anomalous exponent. Note that the unusual
behavior of subdiffusive transport has been observed in an infinite system
with two different values of anomalous exponents \cite{Chechkin2005,*Korabel2010}.

To rectify the structural instability involving unlimited growth of $p(x,t)$%
, at the point of minimum of anomalous exponent $\nu(x)$, we need a
regularization of the fractional equations. The standard approach to
regularize the fractional subdiffusive equations is to temper the power law
waiting time distribution in a such way that the normal diffusion behavior
in the long-time limit is recovered (see, for example, \cite{Meerschaert2008,*Piryatinska2005,*Stanislavsky2008,*Gajda2010}. In this case, by suppressing the power law
behavior, the intrinsic characteristics of the anomalous process are lost.
In this Letter we suggest a completely different approach, where we do not
change the anomalous character, and retain the characteristics of the
process. The main idea is to employ random death process and to `kill' aging
particles, for which the escape rate from the traps tends to zero as age
tends to infinity.

The main aim of this Letter is to show that as long a
death process is introduced, together with a particle production, the
stationary solution of the modified fractional master equation is
structurally stable whatever the spatial variations of anomalous exponent
might be. In particular we use a regularized fractional master equation for
the problem of morphogen gradient formation, which is a central topic of
pattern formation in developmental biology \cite{Hornung}. Here we deal with
a discrete fractional master equation and its continuous approximation,
corresponding to a fractional Fokker-Planck equation.

Let us consider a random walk of particles on a semi-infinite lattice with
unit length. The particle performs a random walk as follows: it waits for a
random time $T_{k}$ at each point $k$ before making a jump to the right with
probability $r(k)$ and left with the probability $l(k)$. We denote the
residence time probability density function by $\psi (k,\tau )=\frac{%
\partial }{\partial \tau }\Pr \left\{ T_{k}<\tau \right\} $, and assume it
has the Pareto form
\begin{equation}
\psi (k,\tau )=\frac{\nu (k)\tau _{0}{}^{\nu (k)}}{\left( \tau _{0}+\tau
\right) ^{1+\nu (k)}},  \notag
\end{equation}%
where $\tau _{0}$ is a constant with the unit of time, and $\nu (k)$ is the
spatially dependent anomalous exponent: $0<\nu (k)<1.$ We assume that during
the time interval $(t,t+\Delta t)$ at point $k$ the particle has a chance $%
\theta (k)\Delta t+o(\Delta t)$ of dying, where $\theta (k)$ is the death
rate ($\theta (k)>0$).

We denote by $p(k,t)$ the average number of particles at the point $k$ at
time $t.$ The anomalous subdiffusive master equation with the death process
can be written as
\begin{align}
\frac{\partial p}{\partial t}& =a(k-1)e^{-\theta (k-1)t}\mathcal{D}%
_{t}^{1-\nu (k-1)}[p(k-1,t)e^{\theta (k-1)t}]  \notag \\
& \quad +b(k+1)e^{-\theta (k+1)t}\mathcal{D}_{t}^{1-\nu
(k+1)}[p(k+1,t)e^{\theta (k+1)t}]  \notag \\
& \qquad \quad -(a(k)+b(k))e^{-\theta (k)t}\mathcal{D}_{t}^{1-\nu
(k)}[p(k,t)e^{\theta (k)t}]  \notag \\
& \qquad \qquad \qquad \qquad \quad \qquad -\theta (k)p(k,t),\quad k\geq 2
\label{eq:master}
\end{align}%
where
\begin{equation}
a(k)=\frac{r(k)}{\Gamma (1-\nu (k))\tau _{0}^{\nu (k)}},\qquad b(k)=\frac{%
l(k)}{\Gamma (1-\nu (k))\tau _{0}^{\nu (k)}}.  \notag
\end{equation}
are the anomalous rate functions. This fractional equation can be derived
from a number of standpoints (see, for example, \cite{mendez2010reaction}).
In this equation the anomalous exponent depends on the state, which is
crucial for what follows. For the case of constant anomalous exponent $\nu $%
, this reaction-transport equation and its continuous approximations were
considered in \cite{Henry2006,Froemberg2008,Yuste2010_2,Fedotov2010}.

To ensure the existence of stationary structure in the long time limit, we
introduce the constant source term $g$ at the boundary of the semi-infinite
lattice ($k=1$). This is crucial for the problem of morphogen gradient
formation, where $g$ models a localized source of morphogens \cite{Yuste2010}%
. We assume that the boundary is reflective, so we have the equation for $%
p(1,t)$
\begin{align}
\frac{\partial p(1,t)}{\partial t}& =b(2)e^{-\theta (2)t}\mathcal{D}%
_{t}^{1-\nu (2)}[p(2,t)e^{\theta (2)t}]  \notag \\
& \qquad -a(1)e^{-\theta (1)t}\mathcal{D}_{t}^{1-\nu (1)}[p(1,t)e^{\theta
(1)t}]  \notag \\
& \qquad \qquad \qquad \qquad \qquad \qquad -\theta p(1,t)+g.
\end{align}%
Note that any nonlinear function $g(p)$ can be included in \eqref{eq:master}.

Without the reaction ( $\theta =0)$ the fractional master equation %
\eqref{eq:master} with constant anomalous exponent $\nu $ is structurally
unstable in the long time limit. The stationary solution $%
p_{st}(k)=\lim_{t\rightarrow \infty }p(k,t)$ can be found from %
\eqref{eq:master} as
\begin{equation}
p_{st}(k)=p_{st}(1)\prod_{j=1}^{k-2}\frac{a(j)}{b(j+1)},\qquad k\geq 2
\end{equation}%
where
\begin{equation}
p_{st}(1)=\left( 1+\sum_{k=2}^{\infty }\prod_{j=1}^{k-1}\frac{a(j)}{b(j+1)}%
\right) ^{-1},
\end{equation}%
provided the sum is convergent. However, when the anomalous exponent is not
constant, the asymptotic behavior is completely different. Consider the
point $M$, at which the anomalous exponent is at a minimum $\nu (M)<\nu (k),$
$\forall k\neq M$. Then, one can show \cite{Fedotov2012} that
\begin{equation}
p(M,t)\rightarrow 1,\qquad p(k,t)\rightarrow 0,\qquad t\rightarrow \infty .
\end{equation}

As stated earlier, the main aim of this Letter is to regularize the
fractional Master equation with the addition of the random death process. To
this end, it is convenient to rewrite the fractional master equation as
\begin{equation}
\frac{\partial p(k,t)}{\partial t}=-I(k,t)+I(k-1,t)-\theta (k)p(k,t),\qquad
k\geq 2  \label{eq:masterI}
\end{equation}%
where $I(k,t)$ is the total flux of cells from $k$ to $k+1$
\begin{align}
I(k,t)& =a(k)e^{-\theta (k)t}\mathcal{D}_{t}^{1-\nu (k)}[p(k,t)e^{\theta
(k)t}]  \notag \\
& -b(k+1)e^{-\theta (k+1)t}\mathcal{D}_{t}^{1-\nu (k+1)}[p(k+1,t)e^{\theta
(k+1)t}].  \label{flux}
\end{align}%
The flux $I(k,t)$, in Laplace space takes the form
\begin{align}
& I(k,s)=a(k)(s+\theta (k))^{1-\nu (k)}\hat{p}(k,s)  \notag \\
& \qquad \qquad -b(k+1)(s+\theta (k+1))^{1-\nu (k+1)}\hat{p}(k+1,s).
\end{align}
From here we can find the stationary flux $I_{st}(k)=\lim_{s%
\rightarrow 0}sI(k,s)$ as follows%
\begin{equation}
I_{st}(k)=a^{\ast }(k)p_{st}(k)-b^{\ast }(k+1)p_{st}(k+1),  \notag
\end{equation}%
where %
\begin{equation}
a^{\ast }(k)=a(k)\left[ \theta (k)\right] ^{1-\nu (k)},\quad
b^{\ast
}(k)=b(k)\left[ \theta (k)\right] ^{1-\nu (k)}  \notag
\end{equation}%
and $p_{st}(k)=\lim_{s\rightarrow 0}sp(k,s)$. The main
feature of this stationary flux is that it has Markovian form; but the rate
functions $a^{\ast }(k)$ and $b^{\ast }(k)$ depend on
the anomalous rate $a(k)$, $b(k)$, random death rate $%
\theta (k)$, and anomalous exponent $\nu (k)$. This unusual
form of stationary flux is because of the non-Markovian character of
subdiffusion.

Let us find the stationary distribution $p_{st}(k)$ for the simple case
where $\theta $ is constant. In the long time limit, at the boundary, we
then have the following condition:
\begin{equation}  \label{eq:Ist1}
I_{st}(1)=g-\theta p_{st}(1).
\end{equation}%
Similarly, we have the condition at the location $k=2$ $I_{st}(2)=I_{st}(1)-%
\theta p_{st}(2)$. We are able to obtain a general expression for the
stationary flux at location $k$
\begin{equation}
I_{st}(k)=g-\theta \sum_{j=1}^{k}p_{st}(j)  \label{I}
\end{equation}%
This has a very simple physical meaning: that as $t\rightarrow \infty $, $%
I_{st}(k)$ tends to the difference between the production rate and the sum
of death rates at all states from the boundary up to $k$. It is clear that
as $k\rightarrow \infty ,$ the stationary flux $I_{st}(k)\rightarrow 0,$
since in the stationary state $g$ should be equal to total death rate
\begin{equation}
g=\theta \sum_{j=1}^{\infty }p_{st}(j).
\end{equation}
We obtain
\begin{align}
b(k+1)\theta ^{-\nu (k+1)}p_{st}(k+1) &= a(k)\theta ^{-\nu(k)}p_{st}(k)
\notag \\
& \quad-\left(\frac{g}{\theta}-\sum_{j=1}^{k}p_{st}(j)\right)  \notag
\end{align}
This equation allows us to find $p_{st}(k)$
for all $k$. For the symmetrical random walk for which $a(k)=b(k)=a$ and $%
\nu=const,$ we have
\begin{equation}
p_{st}(k+1)=p_{st}(k)-\frac{\theta ^{\nu }}{a}\left( \frac{g}{\theta }%
-\sum_{j=1}^{k}p_{st}(j)\right) .  \label{eq:recurrence}
\end{equation}
Now let us obtain the fractional Fokker-Planck equation with the death
process, as the continuous limit of the master equation \eqref{eq:master}.
We change the variables $k\rightarrow x,$ $k\pm 1\rightarrow x\pm l$ and
take the limit $l\rightarrow 0$ to obtain
\begin{align}
\frac{\partial p}{\partial t} &=-\frac{\partial }{\partial x}\left[l\left(
a(x)-b(x)\right) e^{-\theta(x)t}\mathcal{D}_{t}^{1-\nu(x)}[p(x,t)e^{\theta
(x)t}]\right]  \notag \\
&+\frac{\partial ^{2}}{\partial x^{2}}\left[ \frac{l^{2}}{2}%
\left(a(x)+b(x)\right) e^{-\theta(x)t}\mathcal{D}_{t}^{1-\nu
(x)}[p(x,t)e^{\theta(x)t}]\right]  \notag \\
& \qquad \qquad \qquad \qquad \qquad \qquad \qquad -\theta (x)p(x,t).
\label{eq:FFPER}
\end{align}
From this equation we obtain for $p_{st}(x)=\lim_{t\rightarrow \infty }p(x,t)
$ the stationary advection-diffusion equation
\begin{equation}
-\frac{\partial }{\partial x}\left[ v_{\nu }^{\theta }\left( x\right)
p_{st}(x)\right] +\frac{\partial ^{2}}{\partial x^{2}}\left[ D_{\nu
}^{\theta }\left( x\right) p_{st}(x)\right] = \theta (x)p_{st}(x),  \notag
\end{equation}
where $v_{\nu }^{\theta }\left( x\right) $ is the drift, and $%
D_{\nu}^{\theta }\left( x\right) $ is the generalized diffusion coefficient
defined as
\begin{equation}
v_{\nu }^{\theta }\left( x\right) =\frac{l\left( r(x)-l(x)\right) \left[%
\theta (x)\right] ^{1-\nu (x)}}{\Gamma (1-\nu(x))(\tau _{0}{})^{\nu (x)}},
\notag
\end{equation}
\begin{equation}
D_{\nu}^{\theta}\left( x\right) =\frac{l^{2}\left[ \theta (x)\right]^{1-\nu
(x)}}{2\Gamma (1-\nu(x))(\tau_{0})^{\nu (x)}}.  \notag
\end{equation}
This result means that in the long time limit, subdiffusion with the death
process becomes standard diffusion with nonstandard drift $v_{\nu}^{\theta
}\left( x\right)$ and diffusion coefficient $D_{\nu }^{\theta}(x)$. Both of
them depend on the death rate $\theta(x)$ and the anomalous exponent $\nu(x).
$ This is due to non-Markovian character of subdiffusion. It has been found
in \cite{Froemberg2008} that the non-Markovian behavior of subdiffusion
leads to an effective nonlinear diffusion. Note that the drift term $v_{\nu
}^{\theta }\left( x\right) $ plays an essential role in chemotaxis, since $%
v_{\nu }^{\theta }\left( x\right) \sim \frac{\partial C}{\partial x}$, where
$C$ is the chemotactic substance. Therefore the dependence of chemotactic
term of the degradation rate $\theta $ can be of great importance for the
problem of cell aggregation \cite{Othmer1997,Langlands2010,Fedotov2011}.

Let us consider a random walk with a constant drift $v_{\nu }^{\theta }=-v$,
diffusion $D_{\nu }^{\theta }$, and degradation rate $\theta .$ Then
\begin{equation}
v\frac{\partial p_{st}(x)}{\partial x}+D_{\nu }^{\theta }\frac{%
\partial^{2}p_{st}(x)}{\partial x^{2}}-\theta p_{st}(x)=0.
\end{equation}
The solution is the exponential profile
\begin{equation}
p_{st}(x)=A\exp \left[ -\frac{v+\sqrt{v^{2}+4D_{\nu }^{\theta }\theta }}{%
2D_{\nu }^{\theta }}x\right] ,
\end{equation}
where $A$ can be found from the condition $g=\theta
\int_{0}^{\infty}p_{st}(x)dx:$%
\begin{equation}
A=\frac{g\left( v+\sqrt{v^{2}+4D_{\nu }^{\theta }\theta }\right) }{2\theta
D_{\nu }^{\theta }}.
\end{equation}
When $v_{\nu }^{\theta }=0$, we have a morphogen profile obtained in \cite%
{Yuste2010}:
\begin{equation}
p_{st}(x)=\frac{g}{\sqrt{\theta D_{\nu }^{\theta }}}\exp \left[ -\sqrt{\frac{%
\theta }{D_{\nu }^{\theta }}}x\right] .  \label{eq:symstationary}
\end{equation}
We now simulate the fractional master equation with random death process, %
\eqref{eq:masterI}, using Monte Carlo techniques. Throughout this we let $%
\tau_0=1$, so that this is the unit of time for the simulation; we take $g=1$%
, so that we have a constant birth rate of one particle per unit time. The
first particle begins a random walk at $k=1$, such that at each point $k$
waiting times are distributed as $\psi (k,\tau )=\frac{\nu (k)\tau
_{0}{}^{\nu (k)}}{\left( \tau_{0}+\tau \right) ^{1+\nu(k)}}$, and jump
probabilities to the left and right from each point $k$ are $r(k)$ and $l(k)$
respectively. A particle completes a random walk from when it is produced
until the terminal time $t=T$, or until its random time of death
exponentially distributed as $\psi _{D}(t)=\theta e^{-\theta t}$. This death
rate is equivalent to a spatially invariant, constant death rate $\theta$ in %
\eqref{eq:master}. Also note that unlike the waiting time, the death time is
not renewed when the particle makes a jump. The practical issue of having
particles being produced and dying is dealt with in the following way. The
first particle in the simulation begins at time $t=0$, and completes its
random walk as described above; the second particle begins at $t=1$, because
$\tau_0=1$, and completes its random walk; and so on until time $t=T$.

Firstly let us consider the symmetrical random walk, where $r(k)=l(k)=\frac{1%
}{2}$, $\nu (k)=0.5$, and $\theta =10^{-3}$. The figure FIG.~\ref%
{fig:analytical} shows the corresponding stationary density made up from $%
10^{4}$ realizations of the random walk at time $T=10^{6}$.

\begin{figure}[tbp]
\includegraphics[scale=0.35]{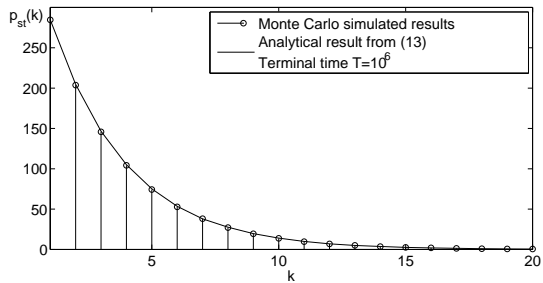}
\caption{Stationary profile for the symmetric fractional master equation
where $r(k)=l(k)=\frac{1}{2}$, $\protect\nu (k)=const.=0.5$, $\protect\tau %
_{0}=1$, and $\protect\theta =10^{-3}$.}
\label{fig:analytical}
\end{figure}
We can see that our simulation is in agreement with the analytical values
calculated from the recurrence relation \eqref{eq:recurrence}.

Next, we show that the model is robust to non-homogenous spatial
perturbations in the anomalous exponent. Analogously to the simulation we
presented in the previous work \cite{Fedotov2012}, we introduce a small
perturbation to the anomalous exponent at one point in the space: all states
have $\nu =0.5$ except for $k=8$, which has $\nu =0.4$. From FIG.~\ref%
{fig:perturb} we can see that although we observe a change to the stationary
profile around point $k=8,$ the stationary profile is structurally stable
and exponential in character. We stress the importance of the death process
in regulating the behavior of the process to ensure stability. Whereas in
our previous work, we showed that even a small perturbation in the anomalous
exponent like this would lead to a breakdown in the stationary density.
Additionally, we considered a non-symmetrical random walk which leads to a
drift and found that the profile is stable.

\begin{figure}[tbp]
\includegraphics[scale=0.35]{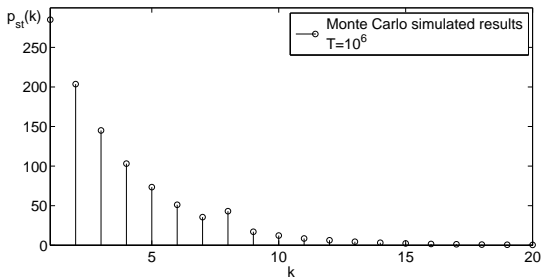}
\caption{Stationary profile for the symmetric fractional master equation,
with a pertubation to the anomalous exponent at $k=8$. $\protect\nu(k \neq
8)=0.5$, $\protect\nu(8)=0.4$}
\label{fig:perturb}
\end{figure}

In summary, we have suggested a new regularization of the subdiffusive
fractional master equation by using the random death process. The
fundamental feature of this approach is that unlike the previous
regularization, which loses the anomalous characteristics, we are able to
retain dependence on the anomalous exponent. We find the stationary flux of
the particles has a Markovian form, with unusual rate function depending on
the anomalous rate functions, the death rate, and the anomalous exponent. We
have shown that the long-time and continuous limit of this regularized
fractional equation is the standard advection-diffusion equation that,
importantly, is structurally stable with respect to spatial variations of
anomalous exponent $\nu$. We have found that the effective advection and
diffusion coefficients, $v_{\nu}^{\theta}$ and $D_{\nu}^{\theta}$, are
increasing functions of the death rate $\theta$: $v_{\nu}^{\theta}\sim
D_{\nu}^{\theta}\sim \theta^{1-\nu}$. We have applied a regularized
fractional master equation and modified fractional Fokker-Planck equation to
the problem of the morphogen gradient formation. We have shown the
robustness of the stationary morphogen distribution against spatial
fluctuations of anomalous exponent.

Acknowledgements:  Sergei Fedotov gratefully acknowledges the support of
EPSRC Grant EP/J019526/1.

\bibliographystyle{apsrev4-1}
\bibliography{ibb}

\end{document}